\documentclass[twocolumn]{aastex62}

\usepackage{natbib}
\shorttitle{The structure of two-sided jets in NGC~1052}
\shortauthors{Nakahara et al.}

\begin{document}

\title{The Two-sided Jet Structures of NGC~1052 at Scales from 300 to $4 \times 10^7$ Schwarzschild Radii}

\author[0000-0001-6894-6597]{Satomi Nakahara}
\affiliation{The Institute of Space and Astronautical Science, Japan Aerospace Exploration Agency, 3-1-1 Yoshinodai, Chuou-ku, Sagamihara, Kanagawa 252-5210, Japan}
\affiliation{Department of Space and Astronautical Science, The Graduate University for Advanced Studies, 3-1-1 Yoshinodai, Chuou-ku, Sagamihara, Kanagawa 252-5210, Japan} 

\author[0000-0003-4384-9568]{Akihiro Doi}
\affiliation{The Institute of Space and Astronautical Science, Japan Aerospace Exploration Agency, 3-1-1 Yoshinodai, Chuou-ku, Sagamihara, Kanagawa 252-5210, Japan}
\affiliation{Department of Space and Astronautical Science, The Graduate University for Advanced Studies, 3-1-1 Yoshinodai, Chuou-ku, Sagamihara, Kanagawa 252-5210, Japan} 

\author{Yasuhiro Murata}
\affiliation{The Institute of Space and Astronautical Science, Japan Aerospace Exploration Agency, 3-1-1 Yoshinodai, Chuou-ku, Sagamihara, Kanagawa 252-5210, Japan}
\affiliation{Department of Space and Astronautical Science, The Graduate University for Advanced Studies, 3-1-1 Yoshinodai, Chuou-ku, Sagamihara, Kanagawa 252-5210, Japan} 

\author[0000-0001-6081-2420]{Masanori Nakamura}
\affiliation{Academia Sinica Institute of Astronomy and Astrophysics, P.O. Box 23-141, Taipei 10617, Taiwan}

\author[0000-0001-6906-772X]{Kazuhiro Hada}
\affiliation{National Astronomical Observatory of Japan, 2-21-1 Osawa, Mitaka, Tokyo 181-8588, Japan}

\author{Keiichi Asada}
\affiliation{Academia Sinica Institute of Astronomy and Astrophysics, P.O. Box 23-141, Taipei 10617, Taiwan}

\author[0000-0001-7719-274X]{Satoko Sawada-Satoh}
\affiliation{Graduate School of Sciences and Technology for Innovation, Yamaguchi University, 1677-1 Yoshida, Yamaguchi-shi, Yamaguchi 753-8512, Japan}

\author[0000-0002-5158-0063]{Seiji Kameno}
\affiliation{Joint ALMA Observatory, Alonso de Cordova 3107 Vitacura, Santiago 763 0355, Chile}
\affiliation{National Astronomical Observatory of Japan, 2-21-1 Osawa, Mitaka, Tokyo 181-8588, Japan}

\correspondingauthor{Satomi Nakahara}
\email{satomi.nakahara@vsop.isas.jaxa.jp}

\received{January 23, 2018} %
\revised{September 5, 2019} 
\accepted{September 11, 2019} 
\submitjournal{Astronomical Journal}

\begin{abstract}
We investigated the jet width profile with distance along the jet in the nearby radio galaxy NGC~1052 at radial distances between $\sim300$ to $4 \times 10^7$ Schwarzschild Radii($R_{\rm S}$) from the central engine on both their approaching and receding jet sides.  
The width of jets was measured in images obtained with the VLBI Space Observatory Programme (VSOP), the Very Long Baseline Array (VLBA), and the Very Large Array (VLA).  The jet-width profile of receding jets are apparently consistent with that of approaching jets throughout the measuring distance ranges, indicating symmetry at least up to the sphere of gravitational influence of the central black hole.    
The power-law index $a$ of the jet-width profile ($w_{\rm{jet}} \propto r^{a}$, where $w_{\rm jet}$ is the jet width, $r$ is the distance from the central engine in the unit of $R_{\rm S}$) apparently shows a transition from $a \sim 0$ to $a \sim 1$, i.e., the cylindrical-to-conical jet structures, at a distance of $\sim1\times10^{4} \ R_{\mathrm{S}}$.   
The cylindrical jet shape at the small distances is reminiscent of the innermost jets in 3C~84.  Both the central engines of NGC~1052 and 3C~84 are surrounded by dense material, part of which is ionized and causes heavy free--free absorption.  
\end{abstract}



\section{INTRODUCTION}\label{section:NGC1052:introduction} 
The structure of astrophysical jet play an important role in accelerating the jet plasma flow in surrounding interstellar/intragalactic medium.  The jet width profile as a function of distance from the central engine in the nearby radio galaxy M87 ($w_{\rm{jet}} \propto z^{a}$; $W_{\rm jet}$ is the jet width, $z$ is the distance from the central engine in the unit of $R_{\rm S}$, and $a$ is the power-law index) exhibits a parabolic structure ($a\sim0.5$) between a few hundred and $10^5$~Schwarzschild radii ($R_{\rm S}$) from the central engine \citep{Asada:2012}.  The region at which the jet acceleration is observed by monitoring observations \citep[e.g.,][]{Nakamura:2013, Mertens:2016} coincides with that of the parabolic streamline \citep{Asada:2014}.  The parabolic structure continues roughly up to the sphere of gravitational influence of the central black hole and is ended with a transition into a conical downstream.  Similar properties with a parabolic-to-conical transition at similar scales (typically $\sim 10^5 R_{\rm S}$) have also been found in NGC~6251 \citep{Tseng:2016} and NGC~4261 \citep{Nakahara:2018}.  These results are thought to be observational evidence that a transverse pressure supports by the external medium with a relatively shallow dependence in the core regions of galaxies and that magnetic energy is dissipated into kinetic energy until they roughly reach comparable as suggested by studies of numerical simulations \citep[e.g.,][]{Komissarov:2009}.  
On the other hand, a parabolic jet structure continues at least up to $\sim 10^8 R_{\rm S}$ without a transition to a conical structure is also suggested for the jets in Cygnus~A \citep{Nakahara:2019}.
Such results can be obtained by measurements over very wide physical scales beyond $\sim 10^5 R_{\rm S}$ by using both the very-long-baseline interferometry~(VLBI) and connected array to cover the angular scales of interest for nearby radio galaxies.  However, the number of examples investigated over such wide scales has been insufficient to examine the diversity of the evolution of jet structure in active galactic nuclei (AGNs).

NGC~1052 is a nearby (a redshift of $0.005037 \pm 0.000020$; \citealt{Denicolo:2005}) radio galaxy whose nuclear activity is classified as low-ionization nuclear emission-line region (LINER).   A well-defined double-sided radio jets elongated by several parsecs with a position angle of approximately 70\degr.  Eastern and Western jets are considered as approaching and receding jets (``jet'' and ``counter jet''), respectively \citep{Kameno:2001,Vermeulen:2003}.  Kpc-scale jet--lobe structures are also prominent on the both sides in the images at 1.4~GHz at arcsec resolutions \citep{Wrobel:1984,Kadler:2004a}.   However, in the VLA image at 5~GHz the central core dominates against weak two-sided radio emissions \citep{Jones:1984}.  NGC~1052 shows a convex radio spectrum peaked at roughly 10~GHz, to be classified into the GHz-Peaked Spectrum~(GPS) radio source \citep{ODea:1998}.  The GPS radiation originates in the central core, which constitutes of pc-scale jets.  
VLBI observations revealed the presence of water masers that distribute along the pc-scale jets.  The water masers were considered to be excited by shocks into circumnuclear molecular clouds, or amplification of the radio continuum emission of the jet by foreground molecular clouds \citep{Claussen:1998}.  On the other hand, the distribution of free--free absorption toward the nuclear region had been investigated in details and suggests the presence of a surrounding obscuring torus with a radius of $\sim 0.5$~pc around the nucleus \citep{Kameno:2001,Kameno:2003}.   The location of large-opacity free--free absorption coincides with the distributions of water masers, HCN, and HCO$^{+}$ molecular absorption lines, almost all of which are redshifted with respect to the systemic velocity of NGC~1052 \citep{Sawada-Satoh:2008,Sawada-Satoh:2016,Sawada-Satoh:2019}.  They argue that these materials around the NGC~1052 nucleus are associated with accreting flows onto the central engine.  
Thus, NGC~1052 is one of well-investigated nearby radio galaxies with apparent two-sided jets.  However, A detailed analysis of the jet
width profile has not previously been published.  

In the present study, we report the measurements of the distance profiles of jet width for NGC~1052.    
Throughout this paper, we adopt a cosmology with $H_0 = 70.5\ h^{-1}$~km~s$^{-1}$~Mpc$^{-1}$, $\Omega_\mathrm{M}= 0.27$, $\Omega_\Lambda= 0.73$ and the redshift corrected to a reference frame defined by the (Virgo + GA + Shapley) model, resulting in an angular size distance of 20.5~Mpc, calculated in the NASA/IPAC Extragalactic Database~(NED).  An angular size of $1$~mas is equivalent to a linear scale of $\sim 0.099$~pc.  Assuming a viewing angle of 86\degr\ \citep{Baczko:2016}, the angular scale $1$~mas is equivalent to $\sim 6,700~R_{\rm S}$, given the mass of the central black hole $10^{8.19}~M_{\rm \odot}$ \citep{Woo:2002}.  The sphere of gravitational influence of the central black hole is estimated to be $\sim 7 \times 10^5 R_{\rm S}$ from a central stellar velocity dispersion of 245~km~s$^{-1}$ \citep[e.g.,][]{Riffel:2017}.  We uncover the structure of jets over a range of $\sim300$ to $4 \times 10^7 R_{\rm S}$ in distance from the central engine of NGC~1052 on both the approaching and receding jets.  
The present paper is structured as follows.  In Section~\ref{section:imageanalysis:NGC1052}, the VLA and VLBI data we used in the present paper and image analysis for jet-width measurements are described.  In Section~\ref{section:NGC1052:results}, we present the results of the jet-width profiles.  In Section~\ref{section:NGC1052:discussion} we make brief discussion of the jet structure and its physical origin.   Finally, our study is summarized in Section~\ref{section:NGC1052:summary}.

\begin{figure}
	\centering
	\includegraphics[width=0.8\linewidth]{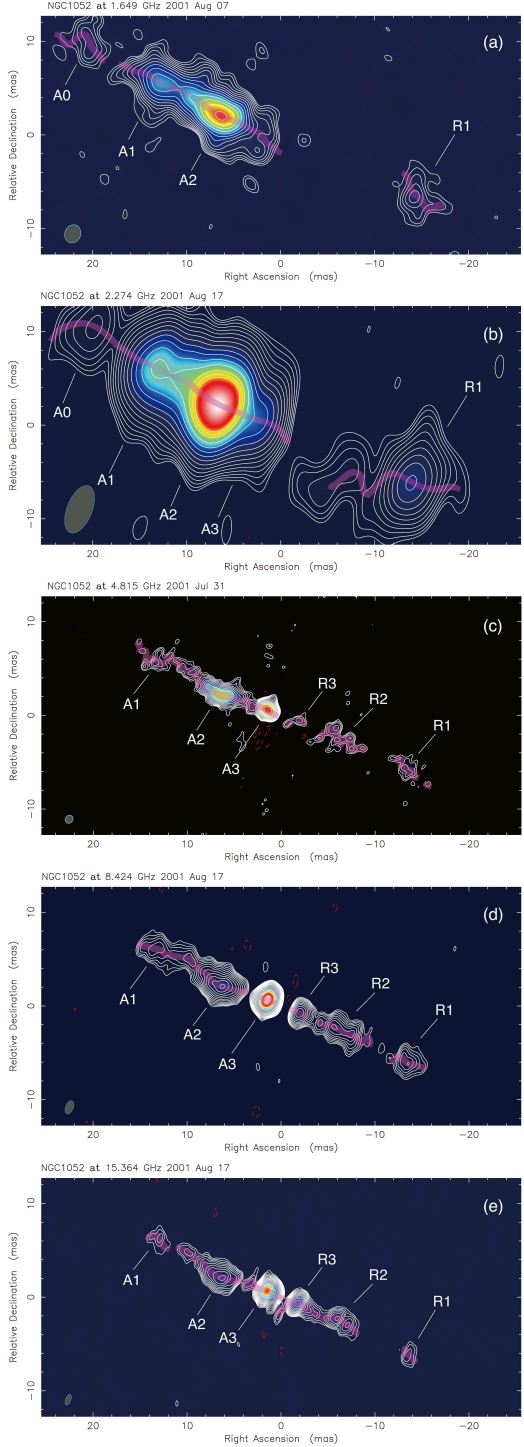}
	\caption{VLBI images of NGC~1052 at 1.6--15.4~GHz on epoch 2001.  (a) VSOP 4.8~GHz, (b) VLBA 2.3~GHz, (c) VSOP 4.8~GHz, (d) VLBA 8.4~GHz, and (e) VLBA 15.4~GHz.  Contours start from three-times the image RMS noise and increase by factors of $\sqrt{2}$.  Convolution was applied by using restoring beam sizes ($\theta_\mathrm{maj}^\mathrm{re} \times \theta_\mathrm{min}^\mathrm{re}$ and $PA^\mathrm{re}$, see Table~\ref{tab:ip1052}).  Semi-transparent curves indicate where the solutions of Gaussian fitting for determining the jet width was successfully obtained.  An angular size of $1$~mas is equivalent to a linear scale of $\sim 0.099$~pc.  Images are centered at the determined black hole position (Section~\ref{section:determineBHposition}).  }
	\label{fig:kameno_images}
\end{figure}
\begin{figure}
	\centering
	\includegraphics[width=\linewidth]{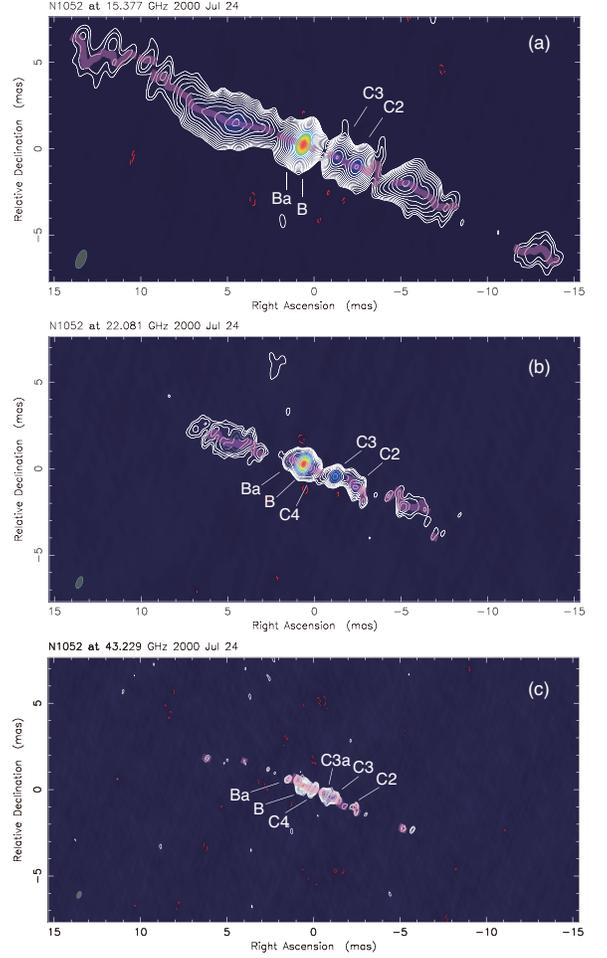}
	\caption{VLBA images of NGC~1052 at 15.4--43.2~GHz on epoch 2000.  (a) VLBA 15.4~GHz, (b) VLBA 22.1~GHz, and (c) VLBA 43.2~GHz.  Contours start from three-times the image RMS noise and increase by factors of $\sqrt{2}$.  Convolution was applied by using restoring beam sizes ($\theta_\mathrm{maj}^\mathrm{re} \times \theta_\mathrm{min}^\mathrm{re}$ and $PA^\mathrm{re}$, see Table~\ref{tab:ip1052}).  Semi-transparent curves indicate where the solutions of Gaussian fitting for determining the jet width was successfully obtained.  An angular size of $1$~mas is equivalent to a linear scale of $\sim 0.099$~pc.  Images are centered at the determined black hole position (Section~\ref{section:determineBHposition}).  }
	\label{fig:sawada_images}
\end{figure}

\section{Data and Image analysis}\label{section:imageanalysis:NGC1052}
\subsection{VSOP data}\label{section:vsop_data}
We used space VLBI images obtained in the VLBI Space Observatory Programme~(VSOP; \citealt{Hirabayashi:1998}) at 1.6~GHz (observation code: W513A) and 4.8~GHz (W513B), which have already been published by \citet{Kameno:2003}.  The observations were carried out on August~7 and July~31, 2001, respectively.  Left-hand circular polarization was received with a bandwidth of 32~MHz.  The spacecraft HALCA was linked to two tracking stations during each observation: Green Bank and Tidbinbilla for W513A, Green Bank and Usuda for W513B.  The ground telescopes were the 10~antennas of the Very Long Baseline Array (VLBA).  In addition to original images, we remade images with uniform weighting and with restoring beams, as listed in Table~\ref{tab:ip1052}.  In preparation for subsequent slice analysis (Section~\ref{section:jetwidthmeasurement:NGC1052}), the restoring beam sizes at the major and minor axes were determined by the projection of the original synthesized beams along perpendicular and parallel to the eastern jet's position angle $PA = 66\degr$, respectively.  The method to define the jet's PA is described in Section~\ref{section:PA}).  The VSOP images at 1.6 and 4.8~GHz shown in Figure~\ref{fig:kameno_images} are the images convolved with the restoring beam; these images have been centered at the black hole position, which will be determined through the manner described in Section~\ref{section:determineBHposition}.

\subsection{VLBA data}\label{section:VLBA_data}
We used multi-frequency images obtained using the VLBA, which are the same data sets that have already been published by \citet{Sawada-Satoh:2008} and \citet{Kameno:2003}.  NGC~1052 was observed using the VLBA at 2.3/8.4/15.4~GHz on 17~August, 2001 \citep[BK084;][]{Kameno:2003}, and at 15.4/22.1/43.2~GHz on 24~July, 2000 \citep[BS080;][]{Sawada-Satoh:2008}.  The details of the observations and data reduction procedures were described in \citet{Sawada-Satoh:2008} and \citet{Kameno:2003}.  Natural weighting images were adopted at 15.4/22.1/43.2~GHz for investigating the jet structures including diffuse emission components; while uniform weighting images were adopted at 22.1/43.2~GHz for investigating the innermost structure of jets by taking advantage of the high spatial resolution.  In addition to original images, we remade images with restoring beams in the same manner as the VSOP cases (Section~\ref{section:vsop_data}).  The VLBA images shown in Figures~\ref{fig:kameno_images} and \ref{fig:sawada_images} are the images convolved with the restoring beams (Table~\ref{tab:ip1052}).

\subsection{VLA data}\label{section:VLA_data}
\begin{figure}
	\centering
	\includegraphics[width=1.0\linewidth]{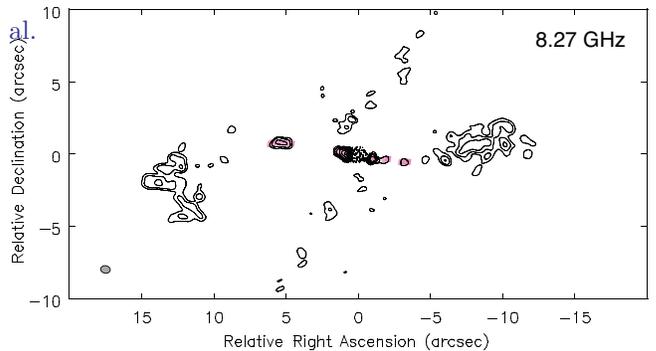}
	\caption{VLA image of NGC~1052 at 8.3~GHz.  A strong point source was subtracted at the map center ($2.57$~Jy).  Contours start from three-times the image RMS noise and increase by factors of $\sqrt{2}$.  Convolution was applied by using restoring beam sizes ($\theta_\mathrm{maj}^\mathrm{re} \times \theta_\mathrm{min}^\mathrm{re}$ and $PA^\mathrm{re}$, see Table~\ref{tab:ip1052}).  Semi-transparent curves indicate where the solutions of Gaussian fitting for determining the jet width was successfully obtained.  An angular size of $1\arcsec$ is equivalent to a linear scale of $\sim 99$~pc.}
	\label{fig:1052-vla-natural}
\end{figure}

The VLA data set AR0396 was retrieved from the data archive system of the National Radio Astronomy Observatory~(NRAO).  We chose the data that has the longest on-source time (33.85~ksec) on NGC~1052 in the VLA archive.  The observation was conducted on 20~June, 1998 at 8.270~GHz using the VLA A-array configuration.  
Dual-circular polarization was received, with a bandwidth of 25~MHz.  The data reduction procedures follow standard manners for VLA data using {\tt AIPS} (Astronomical Image Processing System).  Data of the antennae 3, 7, 9, 19, and 27 were flagged because their system noises were apparently strange.  We performed CLEAN deconvolution using {\tt Difmap} software for image construction.  The image was made with natural weighting and a restoring beam, whose sizes at major and minor axes were defined by projection of the original synthesized beam along directions parallel and perpendicular to jet's position angle $PA = 10\degr$ in the arcsec scale, respectively.       
The image with the restoring beam is shown in Figure~\ref{fig:1052-vla-natural}.  

\subsection{Determination of the central engine location}\label{section:determineBHposition}
The central engine must be located at the upstream end of each side of jet: the eastern/western jet is considered as the approaching/receding jet (or jet/counter jet).    
Radio emission from the innermost nuclear region of NGC~1052 suffers strong free--free absorption (FFA; \citealt{Kameno:2001, Kameno:2003, Kadler:2004,Sawada-Satoh:2008}).  \citet{Kameno:2003} proposed that a subpc-scale circumnuclear torus consisting of dense plasma gas is responsible for observed convex radio continuum spectra due to FFA.  Therefore, there is frequency dependence in the apparent position of the upstream end in the jet.  To determine the position of the central engine, it is necessary to measure the core shift, which is the frequency-dependent position shifts of an observed core due to opacity effect.  However, the core shift for the NGC~1052 jets were determined with some ambiguity \citep{Kadler:2004}, unlike the core shift only due to synchrotron self-absorption in a jet \citep{Lobanov:1998}.  It may be suggesting that the geometry and spatial profiles of dense obscuring materials is not simple (as a similar case, see \citealt{Haga:2015} for NGC~4261).   %

\begin{table*} 
	\begin{center}
		\caption{Image parameters of NGC~1052}\label{tab:ip1052}
		\begin{tabular}{lccccccc}
			\hline \hline
			Telescope	&	$\nu$	&	Obs. date	&	$\sigma$	&	$\theta_\mathrm{maj} \times \theta_\mathrm{min}$			&	$PA$	&	$\theta_\mathrm{maj}^\mathrm{re} \times \theta_\mathrm{min}^\mathrm{re}$			&	$PA^\mathrm{re}$	\\
	&	(GHz)	&	(yyyy-mm-dd)	&	(mJy/beam)	&	(mas $\times$ mas)			&	(deg)	&	(mas $\times$ mas)			&	(deg)	\\
(1)	&	(2)	&	(3)	&	(4)	&	(5)			&	(6)	&	(7)			&	(8)	\\
\hline
VSOP	& $	1.649	$ &	2001-08-07	& $	0.83	$ & $	1.98	\times	1.68	$ & $	-15.5	$ & $	1.97	\times	1.68	$ & $	-22	$ \\
	& $	4.815	$ &	2001-07-31	& $	1.18	$ & $	0.92	\times	0.78	$ & $	-43.4	$ & $	0.89	\times	0.79	$ & $	-22	$ \\
VLBA	& $	2.274	$ &	2001-08-17	& $	0.49	$ & $	6.34	\times	2.58	$ & $	-5.8	$ & $	5.37	\times	2.66	$ & $	-22	$ \\
	& $	8.424	$ &	2001-08-17	& $	0.56	$ & $	1.86	\times	0.77	$ & $	-2.2	$ & $	1.49	\times	0.81	$ & $	-22	$ \\
	& $	15.377	$ &	2001-08-17	& $	0.52	$ & $	1.40	\times	0.55	$ & $	-7.8	$ & $	1.21	\times	0.56	$ & $	-22	$ \\
	& $	15.377	$ &	2000-07-24	& $	0.37	$ & $	1.33	\times	0.52	$ & $	-6.6	$ & $	1.13	\times	0.53	$ & $	-22	$ \\
	& $	22.081	$ &	2000-07-24	& $	1.25	$ & $	0.86	\times	0.34	$ & $	-6.3	$ & $	0.72	\times	0.34	$ & $	-22	$ \\
	& $	43.229	$ &	2000-07-24	& $	0.91	$ & $	0.46	\times	0.20	$ & $	0.2	$ & $	0.35	\times	0.20	$ & $	-22	$ \\
VLA	& $	8.27	$ &	1998-06-20	& $	0.06	$ & $	694	\times	521	$ & $	88.3	$ & $	680	\times	520	$ & $	80	$ \\
			\hline
		\end{tabular}
	\end{center}
	\tablecomments{Columns are as follows: (1) Telescope; (2) Frequency; (3) Observation date; (4) Image RMS noise; (5) Original beam sizes in major and minor axes; (6) Position angle of the major axis for the original beam; (7) Restored beam sizes in major and minor axes; (8) Position angle of the major axis for the restoring beam.}
\end{table*}

Alternatively, there is a method that the ejection point of the approaching and receding jets on the basis of their proper motions is regarded as the central engine position.  This method uses only simple kinematics.  The present study adopts to determine the location of the central engine in the latter way.  The proper motions in the both sides have intensively been measured at 15~GHz, 15~GHz, and 43~GHz by \citet {Vermeulen:2003, Lister:2013}, and \citet{Baczko:2019}, respectively.  Apparent velocity for each blob shows variation in some levels.  Because our images taken on 2000 and 2001 year (Table~\ref{tab:ip1052}), we averaged the jet velocities reported in table~2 of \citet{Vermeulen:2003}, which were based on VLBI monitoring during 1995--2001; we obtained $0.0781\pm0.0022$ pc~yr$^{-1}$ for approaching jets, $0.0783\pm0.0037$ pc~yr$^{-1}$ for receding jets.  Therefore, the position to divide the separation between paired components in the approaching and receding jet sides into 0.0781:0.0783 is considered to be the putative location of the central engine.

Before estimations for the location of central engine, we superimposed images of different frequencies.  
The positions of jet features were determined using {\tt AIPS} {\tt JMFIT} task on all VLBA and VSOP images.  Identified components were labeled as shown in Figures~\ref{fig:kameno_images} and \ref{fig:sawada_images}.   Subsequently, we made pattern matching among the distributions of determined positions at different frequencies in the same manner presented in \citet{Kameno:2003,Sawada-Satoh:2008}.  Because the data of \citet{Kameno:2003} and \citet{Sawada-Satoh:2008} were obtained at different epochs, jet structures changed.  Hence, we made pattern matching for the multi-frequency images of these two groups separately.   

\begin{figure}
	\centering
	\includegraphics[width=\linewidth]{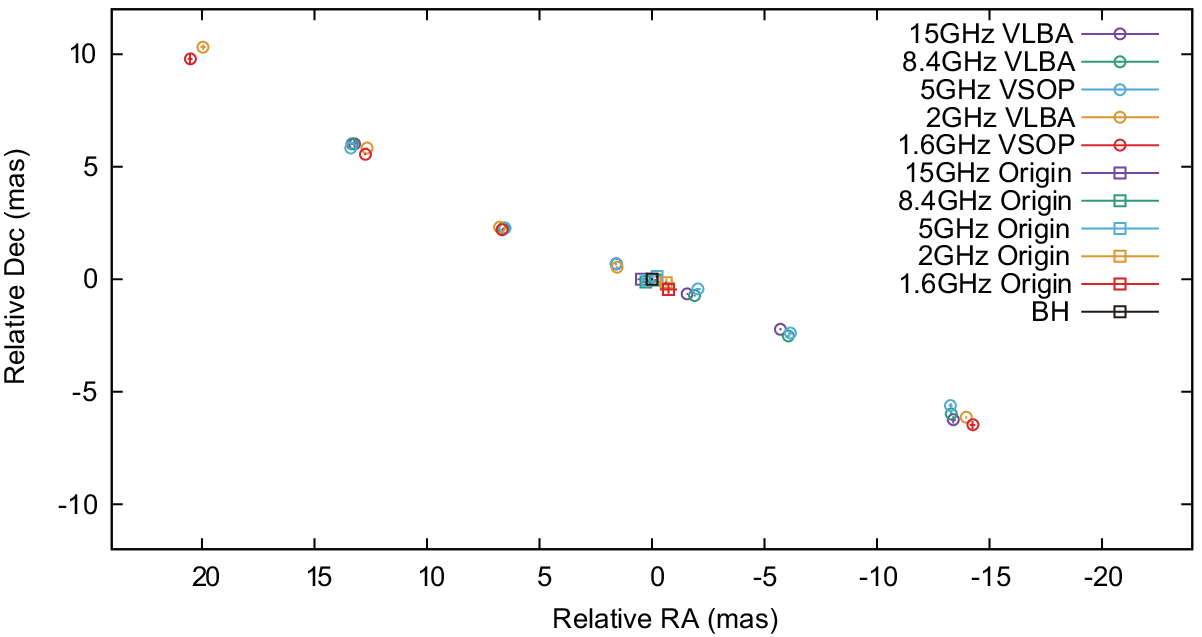}
	\includegraphics[width=\linewidth]{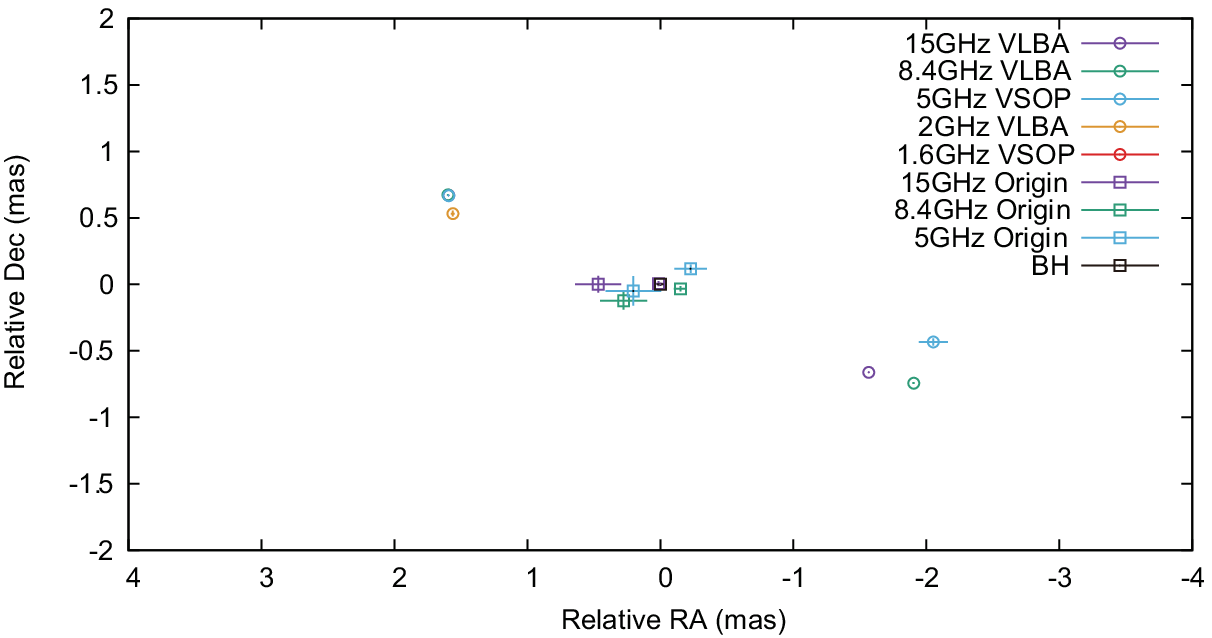}
	\caption{JMFIT results of NGC~1052 at 1.6/2.3/5/8.4/15~GHz for observations on 2001 and position determination of jet origin.  Data sets are the same as \citet{Kameno:2003}.  (Top)~Whole region.  (Bottom)~Focused on central region.  The error bars are the fitting error of JMFIT and the determination error for the jet origins and the black hole (BH).}
	\label{fig:bhposition-1052kameno}
\end{figure}

\begin{figure}
	\centering
	\includegraphics[width=\linewidth]{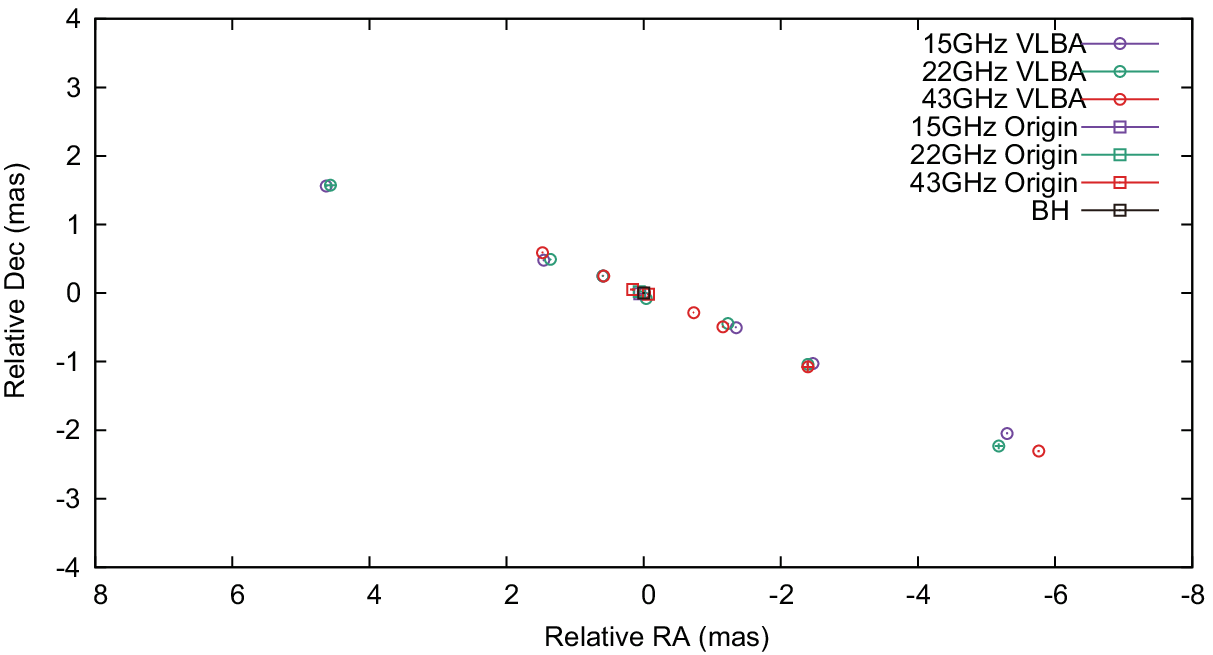}
	\includegraphics[width=\linewidth]{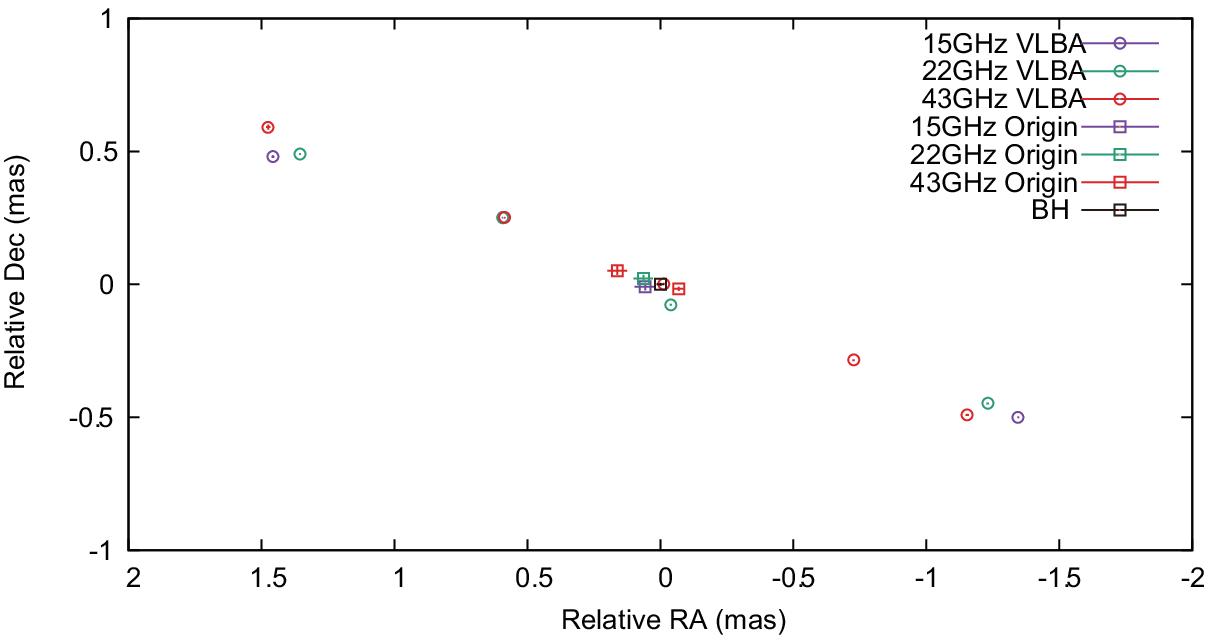}
	\caption{JMFIT results of NGC~1052 at 15/22/43~GHz for observations on 2000 and position determination of jet origin.  Data sets are the same as \citet{Sawada-Satoh:2008}.  (Top)~Whole region.  (Bottom)~Focused on central region.  The error bars are the fitting error of JMFIT and the determination error for the jet origins and the black hole (BH).}
	\label{fig:bhposition-1052sawada}
\end{figure}

Next, we presumed to make pairs of components in the approaching and receding jet sides.  The region where very strong FFA opacities are observed is located at the point between A1--R1 \citep{Kameno:2003} and near the component C4 \citep{Sawada-Satoh:2008}.  Therefore, the defined pairs are A1--R1, A2--R2, A3--R3 for the epoch 2001 summer, B--C3a, A--C1, Ba--C3 for the epoch July~24, 2000 at each frequency in the present study.  
Assuming the apparent jet speeds 0.0781~($\pm 0.0022$)~pc~yr$^{-1}$ and 0.0783~($\pm 0.0037$)~pc~yr$^{-1}$ for the approaching and receding jet sides, respectively, we determined the location of the black hole as the jet production site by dividing the separation of each pair into 0.0781:0.0783.  We obtained the four estimates of the black hole position using the pairs A3--R3, A2--R2, Ba--C3, and B--C3a, which are relatively young jet components.  The four estimates included errors in right ascension and declination, which were propagated from the errors of apparent jet speeds and the image-based {\tt JMFIT} procedure.  Finally, we determined the position of the black hole by weighted means of the two estimates using the A3--R3 and A2--R2 pairs (only at 5, 8.4, and 15 GHz) with uncertainty to 22 $\mu$arcsec for the epoch 2001 summer (Figure~\ref{fig:bhposition-1052kameno}), and of the two estimates using the Ba--C3 and B--C3a pairs with uncertainty to 13 $\mu$arcsec for the epoch July 24, 2000 (Figure~\ref{fig:bhposition-1052sawada}).

As a result, we obtained VLBI images centered at the determined black hole position at all frequencies (Figures~\ref{fig:kameno_images} and \ref{fig:sawada_images}).  The location of the black hole coincides with the intensity peak in the 43~GHz image but is displaced significantly at 22~GHz.  Our result indicates that the line of site to the nucleus becomes optically thin between the two frequencies, which is consistent with conclusions based on intensive VLBI studies at 22--86~GHz for NGC~1052 \citep{Baczko:2016,Baczko:2019}.  
 

\subsection{Determination of the mean jet position angle}\label{section:PA}
In preparation for subsequent slice analysis (Section~\ref{section:jetwidthmeasurement:NGC1052}), we determined weighted-mean PA of the jet axis.  The PAs of the corrected positions of all components, except for innermost components (A3, R3, and C4), on both sides with respect to the black hole (Section~\ref{section:determineBHposition}) were weighted-averaged for each epoch separately.  The propagated errors from the image-based {\tt JMFIT} procedure and the determination of the black hole position (Section~\ref{section:determineBHposition}) was used for the weighting.  We obtained $67\fdg4 \pm 2\fdg3$ and $68\fdg3 \pm 1\fdg6$ from the epochs 2000 and 2001, respectively.  
Therefore, we defined $PA=68\degr$ as the direction to measure angular distance from the black hole in the VLBI scale.  The minor axis of the restoring beams was set to have $PA=68\degr$, while the major one was along $PA=-22\degr$ (Table~\ref{tab:ip1052}).

On the other hand, in the VLA scale, the jet appears toward $PA = 80\degr$ \citep{Kadler:2004a}.   Therefore, we defined $PA=80\degr$ as the direction to measure angular distance from the nucleus at the VLA scale.  The minor axis of the restoring beams was determined along $PA=-10\degr$, while the major one was along $PA=80\degr$ (Table~\ref{tab:ip1052}).

\subsection{Jet width measurements}\label{section:jetwidthmeasurement:NGC1052}
For jet width measurements, we adopted the method similar to those used in the previous studies for NGC~6251 \citep{Tseng:2016} and NGC~4261 \citep{Nakahara:2018}, as described below.  The NGC~1052 jet was assumed to extend at the position angle $PA=68\degr$ (Section~\ref{section:PA}).  We rotated the VLBI images counterclockwise by 22~deg to make the mean jet axis being parallel to the horizontal axis, and then, we defined a new coordinate system ($z_\mathrm{a}$, $r_\mathrm{a}$) centered at the position of black hole.  
$z_\mathrm{a}$ is angular distance from the black hole along the mean jet axis, which is paralleled to the minor axis of the restoring beam.  
$r_\mathrm{a}$ is angular distance from the mean jet axis, which is paralleled to the major axis of the restoring beam.  
The jet intensity profile was transversely sampled, i.e., along $r_\mathrm{a}$-axis, using the {\tt AIPS SLICE} task.  Such slice sampling was performed at different distances $z_\mathrm{a}$.  The pixel sampling rate on both $z_\mathrm{a}$ and $r_\mathrm{a}$ directions was equivalent to approximately fifth of the restoring beam with widths of $\theta_\mathrm{min}^\mathrm{re}$ and $\theta_\mathrm{maj}^\mathrm{re}$, respectively.  We adopted this sampling rate by following previous studies (\citealt{Tseng:2016,Boccardi:2016}, see also a study of angular resolution limit by \citealt{Lobanov:2005}).  
  
The transverse jet profile of each slice were fitted with a Gaussian model to obtain the deconvolved angular size of jet width 
\begin{equation}
w_\mathrm{a} = \sqrt{ (\theta_\mathrm{FWHM})^2 - (\theta_\mathrm{maj}^\mathrm{re})^2 }, \label{eq:decovolution}
\end{equation}
where $\theta_\mathrm{FWHM}$ is the full-width at half maximum~(FWHM) of fitted Gaussian profile and $\theta_\mathrm{maj}^\mathrm{re}$ is equivalent to angular resolution along the jet transverse direction.  
The fitting error of $\theta_\mathrm{FWHM}$ was also obtained, which presumably originated in the image noise, the deconvolution error on imaging processes, and the deviation of the slice profile from the Gaussian model.  
We calculated the error of $w_\mathrm{a}$ by considering the error propagation.  The success criteria for the Gaussian fitting were as follows: (1)~the jet brightness was more than 3-times the image noise, (2) $\theta_\mathrm{FWHM} > \theta_\mathrm{maj}^\mathrm{re}$, and (3)~the determined value of $\theta_\mathrm{FWHM}$ did not exceed its fitting error, which were the same as \citet{Nakahara:2018} used.        
As a result, we obtain the jet width profile, $w_\mathrm{a}(z_\mathrm{a})$, as a function of the distance from the black hole.    

The slice analysis on the VLA image was performed in the same manner, but the different position angle $PA=10\degr$ was assumed.  Note that the original beam was rather elongated roughly at the east--west direction.  Hence, the major and minor axes of the restoring beam were swapped in Eq.~(\ref{eq:decovolution}), and the jet width was obtained by deconvolution with $\theta_\mathrm{min}^\mathrm{re}$ for the VLA case.     
The lobe-like emissions seen in the VLA image at large scales ($|z_\mathrm{a}| \ga 6\arcsec$) were excluded for our analysis.

\begin{figure}
	\centering
	\includegraphics[width=\linewidth]{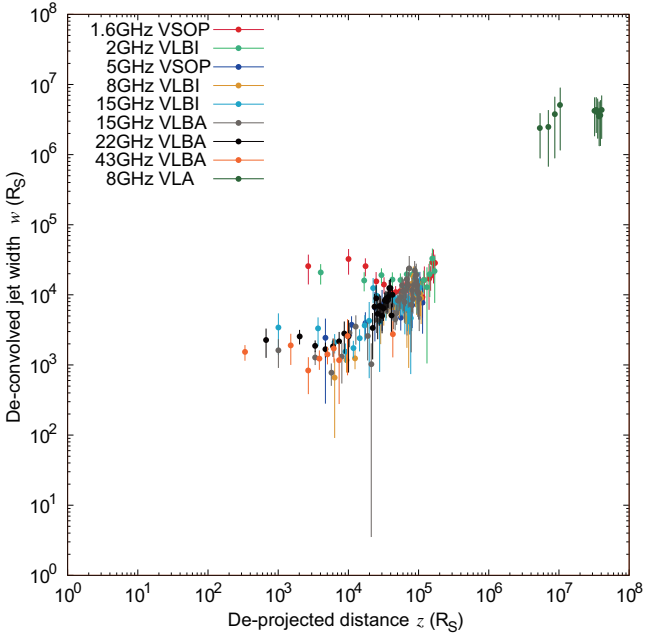}
	\caption{Distance profile of the measured jet width for approaching side of NGC~1052.  The horizontal axis shows the deprojected distance ($z$) from the central engine.  The vertical axis shows the deconvolved jet width ($w$).  The color variation represents observed frequencies; red: 1.6~GHz VSOP, light green: 2~GHz VLBA, blue: 5~GHz for VSOP data, purple: 8~GHz VLBA, light blue: 15~GHz VLBA (Aug~17, 2001), gray: 15~GHz VLBA (Jul~24, 2000), black: 22~GHz VLBA, orange: 43~GHz VLBA, and dark green: 8~GHz VLA, respectively.}
	\label{fig:jetwidth-1052-app}
\end{figure}
\begin{figure}
	\centering
	\includegraphics[width=\linewidth]{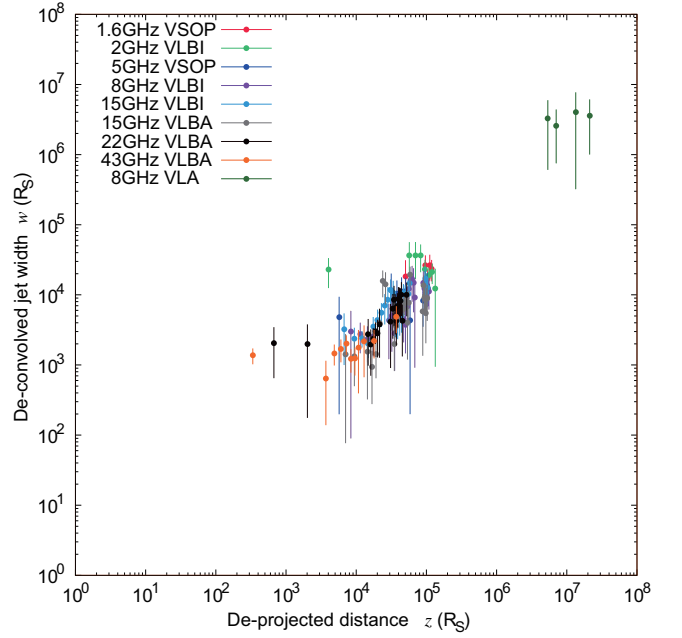}
	\caption{Distance profile of the measured jet width for receding side of NGC~1052.  The horizontal axis shows the deprojected distance ($z$) from the central engine.  The vertical axis shows the deconvolved jet width ($w$).  The color variation represents observed frequencies; red: 1.6~GHz VSOP, light green: 2~GHz VLBA, blue: 5~GHz for VSOP data, purple: 8~GHz VLBA, light blue: 15~GHz VLBA (Aug~17, 2001), gray: 15~GHz VLBA (Jul~24, 2000), black: 22~GHz VLBA, orange: 43~GHz VLBA, and dark green: 8~GHz VLA, respectively.}
	\label{fig:jetwidth-1052-coun}
\end{figure}

\section{RESULTS}\label{section:NGC1052:results}
\subsection{Jet Width Profile for NGC~1052}\label{section:results:NGC1052}
\begin{figure}
	\centering
	\includegraphics[width=\linewidth]{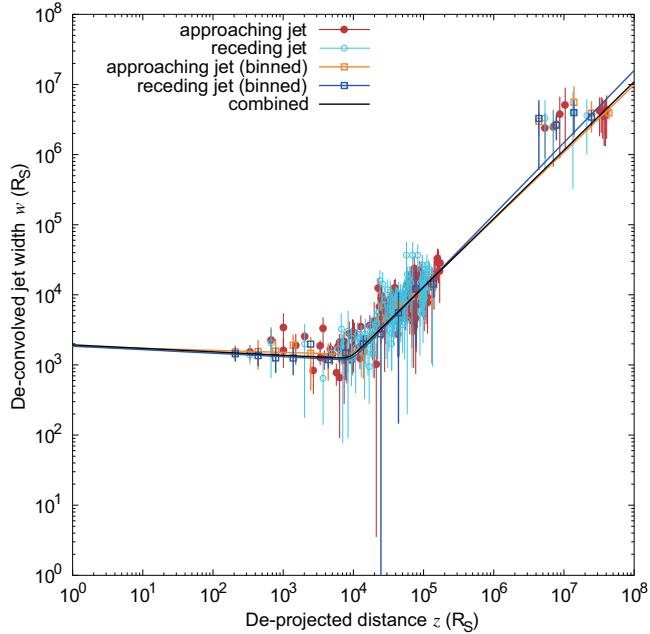}
	\caption{Comparison between approaching jet and receding jet of NGC~1052.  The horizontal axis shows the deprojected distance ($z$) from the central engine.  The vertical axis shows the deconvolved jet width ($w$).  Red and cyan circles represent measurements for approaching and receding jet, respectively.  Orange and blue squares represent the distance binning of the measurements for approaching jet and receding jet, respectively.  Orange, blue, and black solid lines are broken power-law fits for the approaching jet, receding jets, and combined data (approaching + receding), respectively.  
The parameters of fitting results are listed in Table~\ref{table:results-param1-1052}.  
	\label{fig:jetwidth-1052}
	}
\end{figure}

The distance profiles of jet width are shown in Figure~\ref{fig:jetwidth-1052-app} for the approaching jet side and Figure~\ref{fig:jetwidth-1052-coun} for the receding jet side.  Figure~\ref{fig:jetwidth-1052} shows the comparison of them.  
The jet width $w_\mathrm{a}$ in the angular size was converted into $w$ in units of Schwarzschild radii ($R_{\rm S}$).  The angular distance $z_\mathrm{a}$ was converted into the deprojected distance $z$ in units of $R_{\rm S}$ assuming the viewing angle of 86\degr.   
The measurement range in distance resulted in $z_\mathrm{a} = 0.05$~mas--$6.1\arcsec$, corresponding to $z = 0.005$--$610$~pc or $3.3\times10^2$--$4.1\times10^7R_{\rm S}$ from the black hole.     
As a result, our measurements distribute beyond the sphere of gravitational influence $r_\mathrm{SGI} = 7 \times 10^5 R_{\rm S}$ for the black hole of NGC~1052.

At first, the measurements of jet width along the jet had been fitted with a single power-law model function ($w \propto z^a$, where $a$ is the power-law index).  
Since the density of the data sample is significantly higher in the upstream jet ($\la 10^6~R_{\rm S}$) than in the downstream ($\ga 10^6~R_{\rm S}$) as can be seen in Figures~\ref{fig:jetwidth-1052-app} and \ref{fig:jetwidth-1052-coun}, the upstream jets has a much greater influence on the fit.    
Therefore, we binned all the multi-frequency data at intervals of the order of $10^{0.25}$.  
Before the binning, we excluded data of $z \le 10^{4.75} R_{\rm S}$ ($\sim8.4$~mas) at 1.6/2.3/4.8~GHz because a significant deviation from a trend is seen.  We consider the these deviations only at low frequencies is possibly caused by interstellar broadening effect at our line of sight to the innermost jets, which will be studied in a separate paper.      
The single power-law fitting resulted in the power-law indices $a=0.79\pm0.06$ and $a=1.07\pm0.04$, for the approaching side and the receding side, respectively.  The reduced $\chi^2/ndf$ was $45.5/15$ ($p$-value$<10^{-4}$) and $46.7/14$ ($p$-value$<10^{-4}$), where $ndf$ is the number of the degree of freedom indicating that the single power-law does not fit the data well.  

Next, we made the fit with a broken power-law model as the same manner in the previous study for NGC~4261 \citep{Nakahara:2018}.  The fitting function is 
\begin{equation}\label{eq:NGC1052:broken-power-law:width}
w(z)   =   W_0\     2^{(a_{\rm u} - a_{\rm d})/s} \left(\frac{z}{z_{\rm b}}\right)^{a_{\rm u}} \left(1+\left(\frac{z}{z_{\rm b}}\right)^{s}\right)^{(a_{\rm d}-a_{\rm u})/s}, 
\end{equation}
where $W_0$ is the scale factor, 
$a_{\rm u}$ is the power-law index for the upstream jet, 
$a_{\rm d}$ is the power-law index for the downstream jet,
$z_{\rm b}$ is the break location, and 
$s$ is the sharpness of the profile at the break (here, we use the fixed value $s=10$).  
The fitting results are represented in Table~\ref{table:results-param1-1052} and the reduced $\chi^2$ was improved from single power-law fitting.  

\begin{table*}
	\caption{Results of Broken Power-law Fit to Jet Structures of NGC~1052.}
	\begin{center}
		\begin{footnotesize}
		
			\begin{tabular}{lccccccc}
				\hline\hline
				Case	&	$W_0$			&	$z_\mathrm{b}$			&	$a_\mathrm{u}$			&	$a_\mathrm{d}$			&	$\chi^2/ndf$	&	$ndf$	&	$p$-value	 \\
				 & 	($\times 10^3\ R_\mathrm{S}$)			&	($\times 10^4\ R_\mathrm{S}$)			&				&				&		&  & \\
				(1)	&	(2)			&	(3)			&	(4)			&	(5)			&	(6)	&(7) &(8)\\ 
				\hline
				Approaching jet	& $	1.5	\pm	0.5	$ & $	0.99	\pm	0.49	$ & $	-0.03\pm	0.12	$ & $	0.96	\pm	0.04	$ & $	1.10	$ &$13$&$0.35$\\
				Receding jet	& $	1.3 \pm 0.2	$ & $ 0.96 \pm	0.23	$ & $ -0.05 \pm	0.07	$ & $ 1.02 \pm	0.03	$ & $ 0.43 $ &$12$&$0.95$\\
				Approaching + Receding jets	& $	1.3 \pm 0.2 $ & $	0.90	\pm	0.26	$ & $ -0.05 \pm	0.07	$ & $	0.98	\pm	0.03	$ & $	0.75	$&$29$&$0.83$ \\
				\hline
			\end{tabular}
		\end{footnotesize}
		\end{center}
		
\tablecomments{Col.~(1) Cases of fitting analyses; Col.~(2) jet width at structural transition; Col.~(3) distance of structural transition; Col.~(4) power-law index of jet structure at upstream; Col.~(5) power-law index of jet structure at downstream; Col.~(6) reduced chi-squared value; Col.~(7) the number of degree-of-freedom; Col.~(8) probability in the $\chi^2$ distribution.\label{table:results-param1-1052}}
\end{table*}

\section{DISCUSSION}\label{section:NGC1052:discussion} 
On the basis of our investigation for the distance profile of jet width along the jet, we found that the approaching and receding jets are symmetric within errors (Table~\ref{table:results-param1-1052}).  Following the case of NGC~4261 \citep{Nakahara:2018}, NGC~1052 is the second example for the confirmation of symmetry in the two-sided jet structure throughout very wide ranges of physical scales.  Maintaining of symmetry suggests that the distributions of interstellar medium in these host galaxies are spherically symmetric or that the jets have developed with its overwhelming power without being affected by interstellar medium.      
Another important result is that both side of the jet structure changes from a cylindrical ($a\sim0$) to a conical ($a\sim1$) structure at $\sim 10^4 R_{\rm S}$ in NGC~1052.  
Consequently, we found a structural transition in the NGC~1052 jets, as well as those seen in M87 \citep{Asada:2012}, NGC~6251 \citep{Tseng:2016} and NGC~4261 \citep{Nakahara:2018} of nearby radio galaxies.

However, the characteristic feature in the NGC 1052 case is that the jet exhibits a cylindrical-to-conical (from $a\sim0$ to $\sim1.0$) structure, not a parabolic-to-conical (from $a\sim0.5$ to $\sim1.0$) structure as seen in the other these AGNs.     
The profile at $<1 \times 10^4 R_{\rm S}$ in NGC~1052 becomes flat, rather than parabolic, from a conical outer profile (Figure~\ref{fig:jetwidth-1052}).  
The physical scale of the transition is significantly less than $7 \times 10^5 R_{\rm S}$ as the sphere of gravitational influence of the central black hole for NGC~1052.   
The flat profile continues at least $\sim1.5$ orders of magnitude of the distance scale.  \citet{Baczko:2016} estimated the jet width to be $\sim 0.2$~mas at distances of $\sim0.2$~mas on the both sides and to be $\lesssim 0.15$~mas at the innermost component at a distance of $\lesssim15\ \mu$arcsec for the NGC~1052 jets, on the basis of the global mm-VLBI array observation at 86~GHz.  These are corresponding to $\sim1 \times 10^3 R_{\rm S}$ at a distances of $\sim1 \times 10^3 R_{\rm S}$ and $\lesssim 1 \times 10^3 R_{\rm S}$ at a distance of $\lesssim 100R_{\rm S}$.  Their estimates are consistent with our result.  A cylindrical structure ($a \sim 0$) is also inferred at the distance of $\la1$~mas (corresponding to $\la 0.7 \times 10^4 R_{\rm S}$) based on size measurements of each jet component on 43~GHz VLBA images of NGC~1052 at 29~epochs \citep{Baczko:2019}.       
Some physical mechanism to form the flat profile in the inner region is required.   

A relatively flat profile had also been reported in the jet-width distance profile of very young radio galaxy 3C~84.  A limb brightening structure is apparent with a nearly cylindrical flow showing power-law indices of $a=0.25 \pm 0.03$ (at $z\sim180$--$4000 R_{\rm S}$; \citealt{Giovannini:2018}) and $a=0.17 \pm 0.01$ (at $z\sim 10^4$--$10^5 R_{\rm S}$; \citealt{Nagai:2014}) in those researches on the basis of a {\it RadioAstron} and VLBA 43-GHz observations, respectively.    
If the magnetized jets are collimated by the pressure of external medium as theoretical studies suggest, the pressure gradient of ambient gas is required to be $b < 2$, where the pressure profile $p_\mathrm{ism} \propto r^{-b}$ \citep{Komissarov:2009,Porth:2015}, to make an over-collimation with the power-law index of the jet-width profile $a<0.5$.  
\citet{Nagai:2014} discussed that the over-collimation in 3C~84 could be responsible for the ambient pressure gradient shallower.  Alternatively, it may be controlled by different physics because the jet might not have reached pressure balance with the surrounding medium for this very recent outburst ($\sim10$~yr).

The common property of NGC~1052 and 3C~84 as AGNs is that these nuclei are surrounded by the pc-scale distribution of dense obscuring material, part of which is ionized and causes heavy free--free absorption.  For 3C~84, non-detection of receding jet at low frequencies cannot be explained entirely by the Doppler debeaming effect, and therefore, free-free absorption due to the ionized gas is inferred \citep{Walker:2000}.  For NGC~1052, \citet{Kameno:2001} found the presence of a torus-like ionized absorber around the nucleus in addition to molecular gas at outer radii with very high column densities \citep{Sawada-Satoh:2016,Sawada-Satoh:2019}.  It has been argued that these distributions suggest a transition from spherical radial accretion flow to rotational accretion orbits \citep[see also,][]{Kameno:2003,Kameno:2005,Sawada-Satoh:2008}.   
The physical size of the high density ionized region in the center of NGC~1052 is estimated to be $\sim0.5$~pc in radius \citep{Kameno:2001}, corresponding to $\sim3 \times 10^4 R_{\rm S}$; this is consistent with the locations of transition at $\sim1 \times 10^4 R_{\rm S}$ in the jet-width profiles.    
Hence, we consider that the external pressure by the obscuring material is attributed to the observed over-collimation making cylindrical shapes in the innermost jets of NGC~1052, as well as 3C~84.   
After escaping the area of the dense obscuring material, at the outer region, the jets would become conically expanding in surrounding ambient medium with a much steeper profile in the host galaxy and/or become over-pressured by sufficient magnetic dissipation.      
More detailed studies are required to understand the observed jet width profile by numerical simulation including the relationship between a jet and the distribution a circumnuclear torus \citep[e.g.,][]{Fromm:2017}.

\section{Summary}\label{section:NGC1052:summary}
We found the symmetric distance profiles of jet width along the jet throughout scales from 300 to $4 \times 10^7 R_\mathrm{S}$ on the approaching and receding jet sides in the nearby radio galaxy NGC~1052.  The jets show cylindrical-to-conical structures with transitions at $\sim 10^4 R_\mathrm{S}$.  The distance of the transitions is consistent with the subpc-scale distribution of dense obscuring material as a free--free absorber surrounding the central region of NGC~1052.  We suggest a picture that the transverse pressure by the accreting matter in the dense obscuring material supports the over-collimation of the innermost jets in NGC~1052.

\acknowledgments
We are grateful to the anonymous referee for offering constructive comments, which have contributed to substantially improving this paper.  
The VLBA and VLA are operated by the National Radio Astronomy Observatory, which is a facility of the National Science Foundation operated under cooperative agreement by Associated Universities, Inc.  The authors gratefully acknowledge the VSOP Project, which is led by the Japanese Institute of Space and Astronautical Science in cooperation with many organizations and radio telescopes around the world.  This research has made use of the NASA/IPAC Extragalactic Database~(NED) which is operated by the Jet Propulsion Laboratory, California Institute of Technology, under contract with the National Aeronautics and Space Administration.  
This study was partially supported by the Sasakawa Scientific Research Grant from The Japan Science Society (SN) and by Grants-in-Aid for Scientific Research (B) (24340042, AD) from the Japan Society for the Promotion of Science (JSPS).

\end{document}